\documentclass[twocolumn,prb]{revtex4}%
\usepackage{amsfonts}
\usepackage{amsmath}
\usepackage{amssymb}
\usepackage{graphicx}%
\begin{document}
\preprint{ }
\title{Antiferromagnetic Heisenberg model on anisotropic triangular lattice in the
presence of magnetic field}
\author{S. Q. Shen}
\affiliation{Department of Physics, the University of Hong Kong, Pokfulam, Hong Kong, China}
\author{F. C. Zhang}
\affiliation{Department of Physics, University of Cincinnati, Cincinnati, Ohio 45221, USA}
\date{July 2, 2002}

\begin{abstract}
We use Schwinger boson mean field theory to study the antiferromagnetic
spin-1/2 Heisenberg model on an anisotropic triangular lattice in the presence
of a uniform external magnetic field. We calculate the field dependence of the
spin incommensurability in the ordered spin spiral phase, and compare the
results to the recent experiments in Cs$_{2}$CuCl$_{4}$ by Coldea et al.
(Phys. Rev. Lett. 86, 1335 (2001)).

\end{abstract}
\maketitle

The ground states of two-dimensional (2D) Heisenberg models continue to be of
great interest.\cite{chakravarty,auerbach} In this paper, we study the
antiferromagnetic (AF) Heisenberg model on an anisotropic triangular-lattice
in the presence of an external magnetic field along the z-axis,
\begin{equation}
H_{S}=\frac{1}{2}\sum_{i,\delta}J_{\delta}\mathbf{S}_{i}\cdot\mathbf{S}%
_{i+\delta}-\mu B\sum_{i}\mathbf{S}_{i}^{z}.\label{model}%
\end{equation}
In the above equation, the summation runs over all the lattice sites $i$ and
their neighboring sites $(i+\delta)$. We consider AF nearest neighbor
spin-spin couplings, represented by $J$ and $J^{\prime}$ as shown in Fig. 1,
with both $J$ and $J^{\prime}\geq0$. In the absence of the field, this model
is equivalent to a class of models recently considered by a number of
authors.\cite{Zheng99,Merino99,Manuel99,Chung01} The model includes several
well known limiting cases. At $J=0$, it is equivalent to 2D square lattice
model, whose ground state is a two-sublattice N\'{e}el phase. At $J^{\prime
}=0$, it becomes a set of decoupled spin chains. At $J^{\prime}=J$ , it is
reduced to the isotropic triangular-lattice model, where the ground state is a
three-sublattice antiferromagnet. Experimentally, this model may be relevant
to the insulating phase of the layered molecular crystals, $\kappa
$-(BEDT-TTF)$_{2}X$,\cite{Mckenzie} and $\theta$-(BEDT-TTF)$_{2}$
RbZn(SCN)$_{4}$,\cite{Schmalian} Our interest on this model is largely
motivated by recent experiments on Cs$_{2}$CuCl$_{4}$. That system is a quasi
2D $S$ =1/2 frustrated Heisenberg antiferromagnet.\cite{Coldea97} Coldea et
al.\cite{Coldea01} have used the neutron scattering to study the ground state
and the dynamics of the system in high magnetic fields. Among the
observations, these authors found that the incommensurate wave vector changes
as the magnetic field increases, and the spiral spin density wave evolves into
a fully saturated state.%
\begin{figure}
[ptb]
\begin{center}
\includegraphics[
height=1.9in,
width=2.4846in
]%
{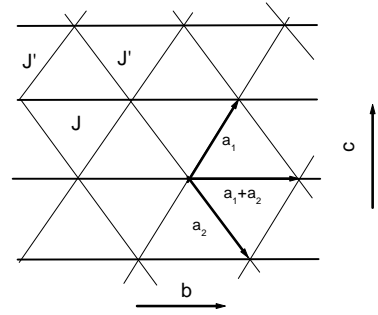}%
\caption{The anisotropic triangle lattice: the bond along the b axis is J and
the bond along the a axis is J'.}%
\end{center}
\end{figure}

In this paper we apply the Schwinger boson mean field theory (MFT) to study
the effect of magnetic field in the frustrated Heisenberg models. This method
enables us to study incommensurate magnetic ordering in the quantum spin
systems. The magnetic ordering is identified as the Bose condensation of the
Schwinger bosons, and the incommensuration of the ordering is determined by
the wave vector of the condensed Schwinger bosons. In the absence of the
magnetic field, the MFT predicts three possible ground state: a two-sublattice
Neel phase, a spiral spin state, and a spin liquid phase, similar to the
results obtained in the high temperature series expansions. \cite{Zheng99} In
the presence of the field, we calculate the field dependence of the
incommensuration in the spiral phase, and compare the results with the
experimental observation in Cs$_{2}$CuCl$_{4}$ with good qualitative agreements.

In terms of Schwinger bosons $a_{i1}$ and $a_{i2}$, the spin operators are
expressed as
\begin{equation}
\mathbf{S}_{i}^{+}=a_{i1}^{\dagger}a_{i2};\mathbf{S}_{i}^{-}=a_{i2}^{\dagger
}a_{i1};\mathbf{S}_{i}^{z}=\frac{1}{2}(a_{i1}^{\dagger}a_{i1}-a_{i2}^{\dagger
}a_{i2})
\end{equation}
with a local constraint at every site $i$ given by $a_{i1}^{\dagger}%
a_{i1}+a_{i2}^{\dagger}a_{i2}=1.$ As a standard method, we introduce a
Lagrangian multiplier field $\lambda_{i}$ to describe the constraint. The
Hamiltonian of the system then becomes
\begin{align}
H  &  =-\frac{1}{8}\sum_{i,\delta,\mu,\nu}J_{\delta}A_{ij,\mu\nu}^{\dagger
}A_{ij,\mu\nu}-\frac{h}{2}\left(  a_{i1}^{\dagger}a_{i1}-a_{i2}^{\dagger
}a_{i2}\right) \nonumber\\
&  +\sum_{i}\lambda_{i}(a_{i1}^{\dagger}a_{i1}+a_{i2}^{\dagger}a_{i2}%
-1)+\frac{1}{8}\sum_{i,\delta}J_{\delta}%
\end{align}
where $h=\mu B;$ $\mu,\nu=1,2$, $j=i+\delta$, and $A_{ij,\mu\nu}=a_{i\mu
}a_{j\nu}-a_{i\nu}a_{j\mu}$ is the spin singlet operator of bond $(ij)$. We
note that operator $A_{ij,\mu\nu}$ is antisymmetric with respect to either the
position $(ij)$ or the indices $(\mu\nu).$ On a bipartite lattice, a spin
rotation by $\pi$ on one of the sublattices transforms the spin-singlet bond
operators into a symmetric operator with respect to the bond indices $(ij)$. A
mean field theory based on that transformation has been developed by Auerbach
and Arovas.\cite{Auerbach88} The method was extended to study the frustrated
lattices by many authors. \cite{Yoshioka91} (For an overview of Schwinger
boson theory, see Ref. [2] and references therein.) We introduce two types of
mean fields: $\Delta_{\mu\nu}(\delta)\equiv\frac{1}{2i}\left\langle
A_{jj+\delta,\mu\nu}\right\rangle $ and $\lambda=\left\langle \lambda
_{j}\right\rangle $ where $\left\langle \cdots\right\rangle $ represents the
thermodynamic average. The mean field Hamiltonian may be solved using the
conventional bosonic Bogliubov transformation as well as the Green function's method.

The mean field Hamiltonian in Eq. (3) is diagonalized as
\begin{equation}
H_{MF}=\sum_{k,\mu=\pm}\omega_{\mu}(k)(\alpha_{k\mu}^{\dagger}\alpha_{k\mu
}+\frac{1}{2})+\mathcal{E}_{0},
\end{equation}
where the single boson spectra $\omega_{\mu}(k)=\omega(k)\pm h/2$ with
$\omega(k)=\sqrt{\lambda^{2}-\left|  \gamma(k)\right|  ^{2}}$ and
$\gamma(k)=\sum_{\delta}J_{\delta}\Delta_{12}(\delta)\sin(k\cdot\delta).$
($\mu=\pm$). $\alpha_{k\mu}$ is a boson annihilation operator related to the
original Schwinger bosons. $\mathcal{E}_{0}/N_{\Lambda}=\sum_{\delta,\mu\nu
}J_{\delta}(\left|  \Delta_{\mu\nu}(\delta)\right|  ^{2}+1/4)/2-2\lambda.$ The
free energy is given by ($\beta= 1/k_{B} T$, with $T$ the temperature)
\begin{equation}
F=\frac{1}{\beta}\sum_{k,\mu=\pm}\ln[1-\exp[-\beta\omega_{\mu}(k)]+\frac{1}%
{2}\sum_{k,\mu=\pm}\omega_{\mu}(k)+\mathcal{E}_{0}.
\end{equation}
The mean field Hamiltonian is solved together with the self-consistent
equations for the two types of mean fields, which are given by
\begin{subequations}
\begin{align}
\frac{1}{N_{\Lambda}}\sum_{k,\mu=\pm}\frac{\lambda}{\omega(k)}\left[
n_{B}(\omega_{\mu}(k))+1/2\right]   &  =2;\label{mf-1}\\
\frac{1}{N_{\Lambda}}\sum_{k,\mu=\pm}\frac{\gamma(k)}{\omega(k)}\sin
(k\cdot\delta)\left[  n_{B}(\omega_{\mu}(k))+1/2\right]   &  =2\Delta(\delta).
\label{mf-2}%
\end{align}
In the MFT, the magnetic ordering may be identified as the Schwinger boson
condensation.\cite{Sarker89} Below we examine the possible Schwinger boson
condensation at the wave vector $k=\pm k^{\ast},$ corresponding to the lowest
energy of $\tilde{\omega}_{\mu}(\pm k^{\ast})\rightarrow0$. We introduce a
non-negative quantity,%

\end{subequations}
\begin{equation}
b_{0}=\frac{2\lambda}{N_{\Lambda}\omega_{-}(k^{\ast})}\left[  n_{B}(\omega
_{-}(+k^{\ast}))+n_{B}(\omega_{-}(-k^{\ast}))\right]  ,
\end{equation}
such that the points $k=\pm k^{\ast}$ in the integrals in Eqs.(\ref{mf-1}) and
(\ref{mf-2}) are taken into account separately,%

\begin{subequations}
\begin{align}
&  \sum_{\mu=\pm}\int\frac{dk^{\prime}}{(2\pi)^{2}}\frac{\lambda}{\omega
(k)}\left[  n_{B}(\omega_{\mu}(k))+1/2\right] \nonumber\\
&  =2-b_{0};\\
&  \sum_{\mu=\pm}\int\frac{dk^{\prime}}{(2\pi)^{2}}\frac{\gamma(k)}{\omega
(k)}\sin(k\cdot\delta)\left[  n_{B}(\omega_{\mu}(k))+1/2\right] \nonumber\\
&  =\frac{2J_{\delta}\Delta(\delta)}{J_{\delta}}-b_{0}\gamma(k^{\ast}%
)\sin(k^{\star}\cdot\delta)/\lambda.
\end{align}
The general features of the mean field solutions in 2D lattices are
qualitatively given as below. At any finite $T$, $b_{0}=0$. At $T=0$, if
$\min[\omega(k)]\neq0$ and $b_{0}=0$, the ground state is a spin liquid, whose
gap depends on the minimal value of the spectra $\omega(k)$. If $\min
[\omega(k)]=0$ and $b_{0}$\ is finite, the Schwinger bosons are condensed to
the lowest energy state, and the system possesses the magnetic long-range
order. The ordering wave vector $Q$ is determined by $\omega(k^{\ast})=0$.
From the spin-spin correlations,
\end{subequations}
\[
\chi_{\alpha\alpha}(q)=-\lim_{\tau\rightarrow0^{-}}\left\langle T_{\tau
}\mathbf{S}_{q}^{\alpha}(\tau)\mathbf{S}_{-q}^{\alpha}(0)\right\rangle \text{
(}\alpha\text{=x,y,z)},
\]
we have
\begin{align}
\chi_{xx}(q)  &  =\chi_{yy}(q)=\chi_{zz}(q)\nonumber\\
&  =\frac{1}{4N_{\Lambda}}\sum_{k}\left(  \frac{n_{B}[\omega(k)]+\frac{1}{2}%
}{\omega(k)}\right)  \left(  \frac{n_{B}[\omega(k+q)]+\frac{1}{2}}%
{\omega(k+q)}\right) \nonumber\\
&  \times\lbrack\lambda^{2}-\gamma(k)\gamma(k+q)]-\frac{1}{16}%
\end{align}
for $h=0,$ which indicates the mean field theory does not break SU(2)
symmetry. In the thermodynamic limit, ($N_{\Lambda}\rightarrow+\infty$), the
correlation functions are convergent except for $Q=2k^{\ast},$%

\begin{equation}
\frac{\chi_{\alpha\alpha}(Q)}{N_{\Lambda}}=\frac{b_{0}^{2}}{8}%
\end{equation}
which indicates that there exists long-range correlations with $Q=2k^{\ast}$
whence $b_{0}\neq0.$

On the triangular-lattice (see Fig. 1), each site has six neighbors:
$\pm\mathbf{a}_{1},\pm\mathbf{a}_{2},\pm(\mathbf{a}_{1}+\mathbf{a}_{2})$. It
is convenient to write the wave vector $k=(k_{1},$ $k_{2})$, with $k_{1}$ and
$k_{2}$ the components of the vector along the directions of $a_{1}$and
$a_{2},$ respectively. The lattice constant $a$ is set to be $1$.

We now consider the mean field solutions at $T=0$. Let me first discuss the
solutions in the absence of the field. At $J^{\prime}/J=1$ (the isotropic
triangular lattice), $k_{0}^{\ast}=\pi/3$ and $\omega(k^{\ast})=0$, $b_{0}$ is
finite indicating a magnetic long range order with the ordering wave vector
$Q(2\pi/3,2\pi/3)$. At $J=0$ (the square lattice), $k_{0}^{\ast}=\pi/2$ and
$\omega(k^{\ast})=0$, and $b_{0}>0$ implying a Neel ordering at $Q(\pi,\pi)$.
The MFT in these two limiting cases is consistent with the known results. At
$J^{\prime}=0$ (decoupled one-dimensional chains), the MFT gives $k_{0}^{\ast
}=\pi/4$, ($2k^{\ast}$ along the chains), $\omega(k^{\ast})>0$, and $b_{0}=0$,
suggesting a spin gap state. The 1D model is exactly soluable, and the ground
state is a gapless spin liquid~\cite{Bethe31}, although the static sin-spin
correlation becomes strongest at $Q_{b}=\pi$.~\cite{Shen98}. The discrepancy
between the MFT and the exact solutions is primarily due to the neglect of the
topological term in the MFT. For the general values of $J^{\prime}/J$, the MFT
predicts three phases at $h=0$: (1). a spin liquid phase at $J^{\prime
}/J<0.136$; (2). a spin spiral state at $0.136<J^{\prime}/J<1.70,$ with the
ordering wave vector bewteen $(\pi/2,\pi/2)$ and $(\pi,\pi)$; (3). an
antiferromagnet with the ordering wave vector $Q=(\pi,\pi)$ at $J^{\prime
}/J>1.70$. We note that the phase diagram of the model was studied previously
by using a series expansion technique, linear spin density wave and SP(N) mean
field theory etc. \cite{Zheng99,Manuel99,Chung01,Yoshioka91} In the method of
series expansion, they found a N\'{e}el state persisting down to $J^{\prime
}/J>1.43$, and predicts a spiral phase at small ratio of $J^{\prime}/J$. These
are consistent with our MFT. These authors also found a dimer phase between
the N\'{e}el and spiral states. The dimer phase breaks translational
invariance and is not included in the present MFT.

We now consider the soltuions at $h\neq0$. In this case, the SU(2) rotational
symmetry is broken, and $\omega_{-}(k)<\omega_{+}(k)$. The Bose condensation
criterion is given by $\omega_{-}(k^{\ast})=\omega(k^{\ast})-h/2=0.$ We use
the MFT to study the field dependent incommensurability in the spin spiral
phase. The results are shown in Fig. 2, where the vector $k^{\ast}$ are
plotted as a function of $J^{\prime}/J$ at several values of $h$. The main
feature is as follows. 1) $J^{\prime}/J=1$\ is a stable fixed point, around
which the external magnetic field does not change the ordering wave vector; 2)
At $0.136<J^{\prime}/J<1$, the ordering wave vector increases slightly as the
field increases; 3) At $J^{\prime}/J>1$, the ordering wave vector decreases as
the field increases; 4) At $J^{\prime}/J<0.136,$ the spin liquid may evolve
into a spiral state, and becomes fully saturated as the field further
increases.%
\begin{figure}
[ptb]
\begin{center}
\includegraphics[
height=1.8455in,
width=2.4111in
]%
{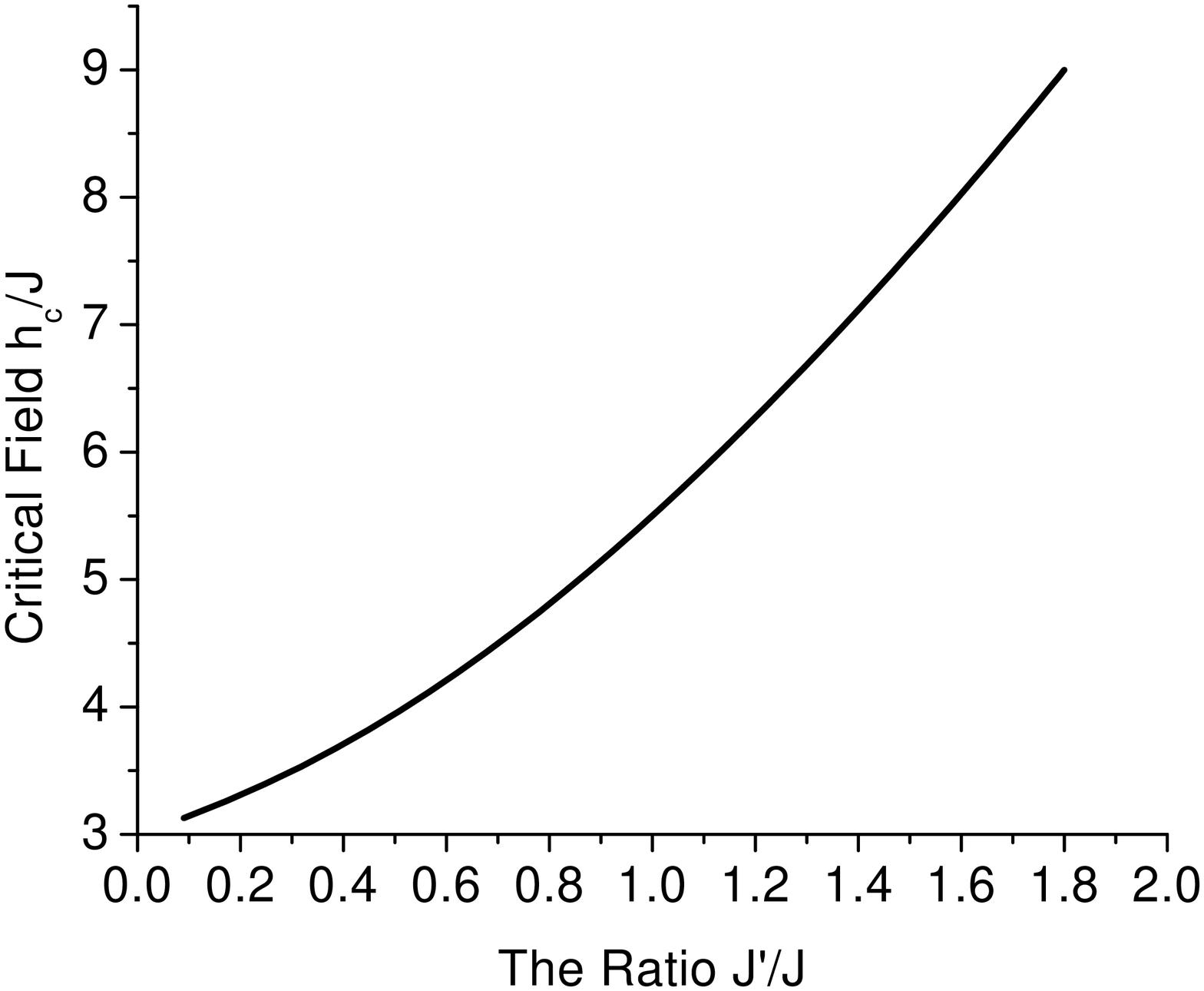}%
\caption{The critical field in unit J versus the ratio $J^{\prime}/J.$}%
\end{center}
\end{figure}

In the presence of the magnetic field, the spin z-component $<S_{z}>\neq0$,
and the ground state breaks the SU(2) invariance. At $T=0$, the expectation
value of $S_{z}$ is given by
\begin{equation}
\left\langle \mathbf{S}_{i}^{z}\right\rangle =\frac{1}{2}h^{\ast}b_{0}%
(k^{\ast},h^{\ast}),
\end{equation}
where $h^{\ast}=h/\left(  2\lambda\right)  $ is the dimensionless field. At
$h\neq0$, $\left\langle \mathbf{S}_{i}^{z}\right\rangle $ is finite, which
indicates a polarized component along the field orientation z-axis. The static
transverse susceptibilities at $Q=2k^{\ast}$ are given by
\begin{equation}
\frac{\chi_{\alpha\alpha}(Q)}{N_{\Lambda}}=\frac{b_{0}^{2}(k^{\ast},h^{\ast}%
)}{8}\left[  1-\left(  h^{\ast}\right)  ^{2}\right]
\end{equation}
for $\alpha=x$ and $y$. $\chi_{\alpha\alpha}(Q)/N_{\Lambda}$ decreases as the
field increases, and approaches to zero at $h^{\ast}=1$. $b_{0}(k^{\ast
},h^{\ast})$ can be calculated within the MFT. A special case is
$b_{0}(k^{\ast},h^{\ast}=1)=1$ corresponding to $\left\langle \mathbf{S}%
_{i}^{z}\right\rangle =1/2$, or the spin full polarization. We have calculated
the critical field $\mu B_{c}$ (defined as the lowest field to induce the full
spin polarization) as a function of $J^{\prime}/J$. The results are plotted in
Fig. 3. As we can see, $\mu B_{c}$ increases as $J^{\prime}$ increases for the
fixed $J$. As $J^{\prime}$ increases, the spin couplings are strengthened, and
it requires a higher field to polarize the spin.%
\begin{figure}
[ptb]
\begin{center}
\includegraphics[
height=1.9726in,
width=2.578in
]%
{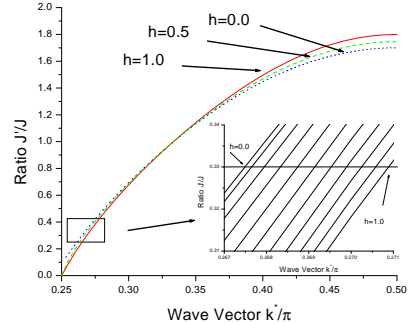}%
\caption{The ratio versus the wave vector $k^{\ast}$ at different fields are
ploted. Inset: The wave vector near $J^{\prime}/J=0.33$ at ten different
fields are ploted. From left to right, $h^{\ast}=0.0,0.1,\cdots,1.0$.}%
\end{center}
\end{figure}

Very recently, Coldea et al.\cite{Coldea01} have reported the neutron
scattering experiments on the antiferromagnet Cs$_{2}$CuCl$_{4}$ in the high
magnetic field. That system is a quasi-2D spin-1/2 quantum system in a
triangular-lattice as shown in Fig.1. The spin-spin couplings are anisotropic
with $J^{\prime}/J\approx0.33$. In the absence of the external magnetic field,
the spins are incommensurately ordered and are aligned within the plane of the
triangular-lattice. The latter may indicate a weak deviation from the
Heisenberg model. Coldea et al. have studied the low temperature states of the
system in the presence of in-plane as well as perperndicular magnetic fields.
In the presence of the perpendicular fields, the states are found magnetically
ordered with the varying incommensuration below a critical field, above which
the system becomes fully spin polarized ferromagnet. In the presence of the
in-plane field, they have observed additional spin liquid phase between the
incommensurate states and the ferromagnetic phase. There have been theoretical
efforts to understand their experimental results.\cite{Bocquet01} ~In the
present paper, we have only considered the Heisenberg model in a uniform
magnetic field. The predicted spin structure breaks SU(2) symmetry, and shows
the spin polarization along the field-direction. Such a spin structure is
compatible with the experiments in the perpendicular field, but incompatible
with the in-plane field. Therefore, our MFT may be of relevance to the
perpendicular field case in their experiments.%
\begin{figure}
[ptb]
\begin{center}
\includegraphics[
height=1.9726in,
width=2.578in
]%
{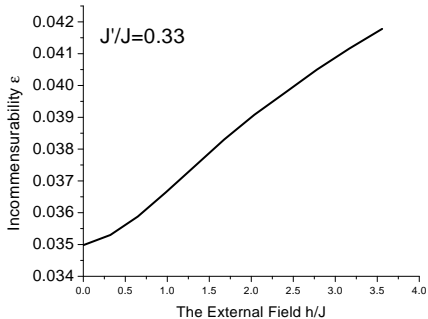}%
\caption{The incommensuration $\epsilon$ versus the external field at
$J^{\prime}/J=0.33$.}%
\end{center}
\end{figure}

To compare with the experiments, we define a quantity to describe the
incommensuration, which is proportional to the wave vector deviation from the
N\'{e}el state, $\epsilon=1/2-Q_{1}/\pi$, with $Q_{1}$ the component of the
spiral wave vector along $a_{1}$ direction. In Table I, we list the
experimental~\cite{Coldea01} and theoretical values of $\epsilon$ at field
$h^{\ast}=0$ and at the critical field $h^{\ast}=1$ for the two values of
$J^{\prime}/J$. Also listed are the values of $\epsilon$ at $h^{\ast}=0$
calculated from the series expansion method~\cite{Zheng99}. The agreements
between the present MFT and the series expansion at $h^{\ast}=0$ are very
good. In Fig. 4, we plot the incommensuration relative to the N\'{e}el state
as a function of the external field for $J^{\prime}/J=0.33$. Our mean field
results are in qualitative agreement with the experiments: as field increases,
$\epsilon$ also increases in the parameter space of interest. Quantitatively,
the theory predicts a weaker variation in the incommensuration than in the
experiments. This discrepancy could be partly due to the neglect of the
deviation of the physical system from the Heisenberg model in the theory.%

\begin{tabular}
[c]{||c|c|c|c|c||}\hline\hline
$h^{\ast}$ & Exp.\cite{Coldea01} & $J^{\prime}/J=0.404$ & $J^{\prime}/J=0.33$
& SE\cite{Zheng99}\\\hline
$0.0$ ($\epsilon_{0}$) & $0.03$ & $0.047$ & $0.035$ & $0.028$\\\hline
$1.0$ ($\epsilon_{c}$) & $0.053$ & $0.053$ & $0.0422$ & --\\\hline\hline
\end{tabular}

Table I: The incommensuration to the N\'{e}el state along the b axis is
listed. $h^{\ast}=0$ means the absence the magnetic field, and $h^{\ast}=1$
means the state is saturated fully. SE means the series expansion method, and
the data is estimated from the work by Zheng et al.\cite{Zheng99}, which was
also predicted by Manuel and Ceccatto.\cite{Manuel99}

In summary, we have used a Schwinger boson mean field theory to study the
antiferromagnetic Heisenberg model on an anisotropic triangular lattice in the
presence of an external magnetic field. We calculate the magnetic field
dependence of the incommensurability of the spin spiral phase. The theoretical
results are compared well with the recent neutron scattering experiments.

We would like to thank stimulating discussions with D. A. Tennant and R.
Coldea on their experiments. This work was supported by a RGC grant of Hong
Kong and a CRCG grant of the University of Hong Kong, and by the US DOE grant
No. FG03-01ER45687, and by the Chinese Academy of Sciences. The authors also
acknowledge ICTP at Trieste, Italy for its support and hospitality, where part
of the work was initiated.

\end{document}